\documentclass[12pt]{article}

\usepackage{amsmath,amssymb}

\global\arraycolsep=1pt
\numberwithin{equation}{section}
\newcommand{\bel}[1]{\begin{equation}\label{#1}}                     
\newcommand{\bal}[1]{\begin{eqnarray}\label{#1}}                     
\newcommand{\be}{\begin{equation}}
\newcommand{\ee}{\end{equation}}

\newcommand{\im}{\mathrm{i}}

\newcommand{\de}{d}

\newcommand{\qq}{\qquad}
\newcommand{\mat}[1]{\begin{pmatrix} #1 \end{pmatrix}}
\newcommand{\ul}[1]{\underline{#1}}
\newcommand{\ol}[1]{\overline{#1}}
\renewcommand{\thefootnote}{\fnsymbol{footnote}}

\begin{document}

\begin{flushright}
January, 2008 \\
OCU-PHYS 288 \\
\end{flushright}
\vspace{20mm}

\begin{center}
{\bf\Large
Nambu-Goto Like Action \\
for the $AdS_5 \times S^5$ Superstrings \\
in the Generalized Light-Cone Gauge\\}
\end{center}

\begin{center}

\vspace{15mm}

Hiroshi Itoyama$^{ab}$\footnote{
\texttt{itoyama@sci.osaka-cu.ac.jp}
}, 
Takeshi Oota$^b$\footnote{
\texttt{toota@sci.osaka-cu.ac.jp}
} and
Reiji Yoshioka$^a$\footnote{
\texttt{yoshioka@sci.osaka-cu.ac.jp}
}

\vspace{10mm}

\textit{${}^a$Department of Mathematics and Physics, 
Graduate School of Science,\\
Osaka City University\\
3-3-138 Sugimoto, Sumiyoshi,
Osaka 558-8585, JAPAN
}
\vspace{5mm}

\textit{${{}^b}$Osaka City University
Advanced Mathematical Institute (OCAMI)\\
3-3-138 Sugimoto, Sumiyoshi,
Osaka 558-8585, JAPAN
}

\vspace{5mm}

\end{center}
\vspace{8mm}

%%%%%%%%%%%%%%%%%%%%%%%%%%%%%%%%%%%%%%%%%%%%%%%%%%%%%%%%%%%%%%%%%%%%%%%
\begin{abstract}
We reinvestigate the $\kappa$-symmetry-fixed Green-Schwarz action 
in the $AdS_5 \times S^5$ background 
in a version of the light-cone gauge. 
In the generalized light-cone gauge,
the action has been written in the phase space variables. 
We convert it into the standard
action written in terms of the fields and their derivatives. 
We obtain a Nambu-Goto type action which has the correct flat-space limit.
\end{abstract}
%%%%%%%%%%%%%%%%%%%%%%%%%%%%%%%%%%%%%%%%%%%%%%%%%%%%%%%%%%%%%%%%%%%%%%%

\vspace{25mm}

\newpage

\renewcommand{\thefootnote}{\arabic{footnote}}
\setcounter{footnote}{0}

%%%%%%%%%%%%%%%%%%%%%%%%%%%%%%%%%%%%%%%%%%%%%%%%%%%%%%%%%%%%%%%%%%%%%%%

\section{Introduction and Summary}

It is a challenging problem to
quantize the Green-Schwarz (GS) action \cite{GS,GS2} 
in the $AdS_5 \times S^5$ background \cite{MT}.
The knowledge of the spectrum will tell us the strong coupling dynamics
of the large $N$ gauge theory through the AdS/CFT correspondence.
One of the difficulties in covariant quantization of
the GS action stems from the existence
of the local $\kappa$-symmetry, 
which halves the fermionic degrees of freedom
\cite{GS,GS2,HK}.
One approach to this problem is to abandon 
the covariance and fix the $\kappa$-symmetry
non-covariantly.
But after the $\kappa$-symmetry fixing, the model is 
still a constrained system due to
the world-sheet diffeomorphism. 
Various gauges \cite{pes,kal,KR98,KT98,MT00,MTT00,CMS,KRT04,
KT04,AF04,AF05,FPZ}
have been proposed to fix these symmetries.
Especially, the uniform light-cone gauge, a generalization of the 
flat-space light-cone gauge \cite{GGRT} to the curved space background,
has been extensively investigated in \cite{CMS,KRT04,KT04,AF04,AF05,FPZ}.

In our previous paper \cite{IO}, 
we adopted the light-cone gauge and considered
the Hamiltonian dynamics of the GS action by using the physical degrees of freedom.
The action is formulated in the first order formalism, i.e.,
is written in terms of the phase space variables.
This first-order formulation is suited for considering the
problem of canonical quantization.

Unfortunately, the reduced action in the generalized light-cone gauge
still has an involved form. Before considering the quantization
problem of the action, we would like to investigate quantum 
fluctuation in various limits. Extensive study around the plane-wave
region was done in \cite{CMS}.
But, in general, the first-order Lagrangian
is not so convenient to study the quantum spectrum in various limits,
such as the BMN limit \cite{BMN}, the Hofman-Maldacena limit \cite{HM}, or
Maldacena-Swanson limit \cite{MS}.
They can be better investigated by using the Lagrangian
written in terms of the fields and their derivatives.
Therefore, in this paper, we reformulate the GS action 
to the standard form
in the generalized light-cone gauge.

After $\kappa$-symmetry fixing and taking the generalized light-cone gauge
\be
X^+ = \kappa \tau, \qq P_- = \sqrt{\lambda} \omega,
\ee
with $\kappa$ and $\omega$ are constants, 
we find the Nambu-Goto like action for the GS action in the $AdS_5 \times S^5$
background. Its bosonic part has the following form:
\be
S = \frac{1}{2\pi} \int \de^2 \xi
\left( \sqrt{ - \lambda \det( \mathcal{J}_{ij} + \mathcal{G}_{ij} )}
+ \sqrt{\lambda} \kappa \omega \frac{G_{+-}}{G_{--}} \right).
\ee
Here $\mathcal{G}_{ij} = G_{mn} \partial_i X^m \partial_j X^n$ is the induced
metric, induced from the target space metric for the transverse spatial directions.
(The indexes $m,n$ run over the transverse directions: $m,n=1,2,\dotsc, 8$).
The additional term $\mathcal{J}_{ij}$ comes from the longitudinal directions.
The 't Hooft coupling $\lambda$ is related to the radius $R$
of $AdS_5$ and $S^5$ as $\sqrt{\lambda} = R^2/\alpha'$.
The full Lagrangian which includes the fermions will be given by \eqref{FLGS}.
This is a Nambu-Goto type Lagrangian.

It is natural to appear the Nambu-Goto type action when the world-sheet diffeomorphism
is fixed by certain gauge conditions other than the conformal gauge. 
Solving the equations
of motion for the world-sheet metric yields the Nambu-Goto type action.
For example, the Nambu-Goto type action for the GS model in $AdS_5 \times S^5$
in the static gauge 
can be found in \cite{KT98}.

In contract to ordinary Nambu-Goto actions, 
we have chosen the sign before the square root term to be positive.
This comes from the requirement that the action must have the correct flat-space limit.  
Indeed, the Lagrangian \eqref{FLGS} goes to the correct $\kappa$-symmetry fixed
light-cone gauge
Lagrangian in the limit.

The Lagrangian \eqref{FLGS} will serves as a starting point for developing the
various limits and investigating the quantum fluctuations.

This paper is organized as follow.
In section 2, using the bosonic sigma model as an example,
we explain the procedure to obtain the standard Lagrangian in the generalized
light-cone gauge. In section 2.1, we start from the Lagrangian in the first-order
formalism and arrive at the standard one. In section 2.2, we derive
it without going to the first-order form.
In section 3, we first briefly review our notation for the GS action.
In section 3.1, the $\kappa$-symmetry fixing is done and the action
in the $AdS_5 \times S^5$ background is given. In section 3.2,
the $\kappa$-symmetry fixed GS action in the generalized light-cone gauge is obtained.
This is our main result. In section 3.3, 
it is discussed that the action has the correct flat-space limit.
Some of our notations are summarized in Appendix.

\section{The bosonic sigma model}

The action for the bosonic sigma model 
in the $D$-dimensional curved target space is given by
\be
S = \frac{1}{2\pi} \int \de^2 \xi \, \mathcal{L},
\ee
where
\bel{BSM}
\mathcal{L} = - \frac{1}{2} \sqrt{\lambda} h^{ij} G_{\ul{m}\, \ul{n}}
\partial_i X^{\ul{m}} \partial_j X^{\ul{n}},
\ee
Here $\ul{m},\ul{n}=0,1,\dotsc, D-1$, $\xi^0 = \tau$, $\xi^1=\sigma$,
$\partial_i = \partial/\partial \xi^i$, 
$h^{ij} = \sqrt{-g} g^{ij}$
$(i,j=0,1)$.

The conjugate momenta are given by
\be
P_{\ul{m}} = - \sqrt{\lambda} h^{0j} G_{\ul{m}\, \ul{n}} 
\partial_j X^{\ul{n}}.
\ee
The target space metric is assumed to have the following form
\bel{TSM}
G_{\ul{m}\, \ul{n}} \de X^{\ul{m}} \de X^{\ul{n}}
= G_{\mathfrak{a}\mathfrak{b}} \de X^{\mathfrak{a}} 
\de X^{\mathfrak{b}}
+ G_{mn} \de X^m \de X^n.
\ee
Here $\mathfrak{a}, \mathfrak{b}= \pm$ denote the longitudinal directions, 
$m,n=1,2,\dotsc, D-2$ denote the transverse directions.
We assume that $\partial/\partial X^{\pm}$ is a Killing vector.

Let us decompose the Lagrangian into two pieces:
\be
\mathcal{L} = \mathcal{L}_1 + \mathcal{L}_2,
\ee
\be
\mathcal{L}_1 = - \frac{1}{2} \sqrt{\lambda} h^{ij}
G_{\mathfrak{a} \mathfrak{b}} \partial_i X^{\mathfrak{a}}
\partial_j X^{\mathfrak{b}},
\ee
\be
\mathcal{L}_2 = - \frac{1}{2} \sqrt{\lambda} h^{ij}
G_{mn} \partial_i X^m \partial_j X^n.
\ee
The first part $\mathcal{L}_1$ and the second part $\mathcal{L}_2$ 
are related to the metric for the longitudinal directions and the metric
for the transverse directions respectively. 

Under the target space metric ansatz \eqref{TSM}, the momenta $P_-$
which is conjugate to $X^-$ is given by
\be
P_- = - \sqrt{\lambda} h^{0j} G_{-,\mathfrak{a}} \partial_j 
X^{\mathfrak{a}}.
\ee
The generalized light-cone gauge is given by the following two conditions\footnote{
Originally, the second condition was given by $\partial_1(P_-) = 0$. Most general
solution is $P_- = P_-(\sigma)$. But without loss of generality,
we can set $P_-$ to a constant by redefining the world-sheet space variable $\sigma$
and conjugate momenta
such that
$P_-(\sigma) \de \sigma = P'_- \de \sigma'$ with $P_-'$ constant. Therefore,
we adopt the condition $P_- = \mathrm{const}$ as one of the gauge conditions.}:
\be
X^+ = \kappa \tau, \qq
P_- = \sqrt{\lambda} \omega = \mathrm{const},
\ee
which fixes the world-sheet diffeomorphism.

\subsection{From the first order form to the standard Lagrangian}

The reduced action in the generalized light-cone gauge 
is given by (we use the notation of \cite{IO})
\bel{1stS}
S_{\mathrm{red}} = \frac{1}{2\pi} \int \de^2 \xi
\bigl( P_m \dot{X}^m - \mathcal{H}_{\mathrm{LC}} \bigr),
\ee
where
\be
\mathcal{H}_{\mathrm{LC}} = - \kappa P_+.
\ee
This is the first-order Lagrangian $\mathcal{L} = \mathcal{L}(X^m, P_m)$
written in terms of the transverse coordinates $X^m$ and their conjugate momenta
$P_m$.

Here $P_+$ is a solution of
\bel{quadPp}
G^{++} P_+^2 + 2 \sqrt{\lambda} \omega G^{+-} P_+ + C = 0,
\ee
where
\bel{C}
C = \lambda \omega^2 G^{--} + \lambda G_{mn} \partial_1 X^m 
\partial_1 X^n
+ K^{mn} P_m P_n,
\ee
\be
K^{mn} := G^{mn} + \frac{1}{\omega^2} G_{--} \partial_1 X^m
\partial_1 X^n.
\ee
Explicitly, $P_+$ is given by
\be
P_+ = \frac{\varepsilon}{G^{++}} 
\sqrt{ \lambda ( G^{+-})^2 \omega^2 - G^{++} C}
- \sqrt{\lambda} \omega \frac{G^{+-}}{G^{++}},
\ee
where $\varepsilon = \pm 1$. 

Now let us convert this first-order Lagrangian into the standard form.
The equations of motion for $P_m$
\be
\dot{X}^m + \kappa \frac{\partial P_+}{\partial P_m} = 0
\ee
yield the following relations
\be
\varepsilon \sqrt{ \lambda \omega^2 (G^{+-})^2 - G^{++}C } \, 
\dot{X}^m = \kappa K^{mn} P_n.
\ee
It is convenient to introduce $\mathcal{J}_{ij}$ and 
$\mathcal{G}_{ij}$ by
\be
\mathcal{J}_{00} := \frac{\kappa^2}{G^{++}}, \qq
\mathcal{J}_{11} := \frac{\omega^2}{G_{--}}, \qq
\mathcal{J}_{01} = \mathcal{J}_{10} := 0,
\ee
\be
\mathcal{G}_{ij} := G_{mn} \partial_i X^m \partial X^n, \qq
i,j=0,1.
\ee
Let $K_{mn}$ be the inverse of $K^{mn}$:
\be
K_{mn} = G_{mn} - \frac{( G_{mm'} \partial_1 X^{m'})
( G_{nn'} \partial_1 X^{n'} ) }{\mathcal{J}_{11} 
+ \mathcal{G}_{11}}.
\ee
Then
\bel{Pm}
P_m = \frac{\varepsilon}{\kappa} 
\sqrt{ \lambda \omega^2 ( G^{+-})^2  - G^{++} C } \, 
K_{mn} \dot{X}^n.
\ee
By substituting this relation into \eqref{C}, we have
\be
C = \lambda \omega^2 G^{--} + \lambda \mathcal{G}_{11}
+ \frac{1}{\kappa^2} \left( \lambda \omega^2 (G^{+-})^2 - G^{++}C
\right) K_{mn} \dot{X}^m \dot{X}^n.
\ee
Then we have
\bel{C2}
C = \frac{\lambda \omega^2 G^{--} + 
\lambda \mathcal{G}_{11} + (\lambda \omega^2/\kappa^2) 
(G^{+-})^2 K_{mn} \dot{X}^m \dot{X}^n }
{1 + (1/\kappa^2) G^{++} K_{mn} \dot{X}^m \dot{X}^n}.
\ee
Note that
\be
K_{mn} \dot{X}^m \dot{X}^n 
= \frac{\mathcal{J}_{11} \mathcal{G}_{00} + \det( \mathcal{G}_{ij})}
{ \mathcal{J}_{11} + \mathcal{G}_{11}}.
\ee
With some work, we find
\be
\lambda \omega^2 ( G^{+-})^2 - G^{++} C
= - \frac{\lambda \kappa^2}{\det( \mathcal{J}_{ij} + 
\mathcal{G}_{ij})}
( \mathcal{J}_{11} + \mathcal{G}_{11})^2.
\ee
Assuming $\det( \mathcal{J}_{ij} + \mathcal{G}_{ij}) < 0$, we have
\be
\sqrt{\lambda \omega^2 ( G^{+-})^2 - G^{++} C}
= \varepsilon' \sqrt{- \frac{\lambda}{\det( \mathcal{J}_{ij} 
+ \mathcal{G}_{ij})}}
\, \kappa ( \mathcal{J}_{11} + \mathcal{G}_{11}).
\ee
Here $\varepsilon' = \mathrm{sign}( \kappa ( \mathcal{J}_{11} 
+ \mathcal{G}_{11}))$.

Let $J_{ij}$ be a matrix defined by
\be
J_{ij}:= \mathcal{J}_{ij} + \mathcal{G}_{ij},
\ee
and $J^{ij}$ be its inverse.
We can see that
\be
K_{mn} \dot{X}^n = \frac{1}{\mathcal{J}_{11} + \mathcal{G}_{11}}
G_{mn} \bigl[ ( \mathcal{J}_{11} + \mathcal{G}_{11}) \dot{X}^n
- \mathcal{G}_{01} \partial_1 X^n \bigr]
= \frac{\det( \mathcal{J}_{ij} + \mathcal{G}_{ij})}
{\mathcal{J}_{11} + \mathcal{G}_{11}} \, G_{mn} J^{0j} \partial_j X^n.
\ee

Now we finally have 
\be
P_m = - \varepsilon \varepsilon' \sqrt{-\lambda \det( J_{ij})}\, 
G_{mn} J^{0j} \partial_j X^n.
\ee
By substituting this expression into the first order form 
of the action,
we get the reduced Lagrangian in the generalized light-cone gauge.

Summary: The Lagrangian of the bosonic sigma model in the generalized
light-cone gauge is given by
\be
S_{\mathrm{red}} = \frac{1}{2\pi} \int \de^2 \xi\, 
\mathcal{L}_{\mathrm{LC}},
\ee
where
\bel{BSM2}
\mathcal{L}_{\mathrm{LC}}
= - \varepsilon \varepsilon' \sqrt{ - \lambda
\det( \mathcal{J}_{ij} + \mathcal{G}_{ij})}\, 
+ \sqrt{\lambda} \kappa \omega \frac{G_{+-}}{G_{--}}.
\ee
Here
\be
\mathcal{J}_{00} = \frac{\kappa^2}{G^{++}}, \qq
\mathcal{J}_{11} = \frac{\omega^2}{G_{--}}, \qq
\mathcal{J}_{01} = \mathcal{J}_{10} = 0,
\qq
\mathcal{G}_{ij} = G_{mn} \partial_i X^m \partial_j X^n.
\ee

\subsection{Rederivation of the reduced Lagrangian}

In this subsection, we rederive the reduced Lagrangian 
\eqref{BSM2} without
bypassing the first-order formalism.

Let us restart from the Lagrangian \eqref{BSM}.
Let us decompose it as follows
\be
\mathcal{L} = \mathcal{L}_1 + \mathcal{L}_2,
\ee
\be
\mathcal{L}_1 = - \frac{1}{2} \sqrt{\lambda} h^{ij}
G_{\mathfrak{a} \mathfrak{b}}
\partial_i X^{\mathfrak{a}} \partial_j X^{\mathfrak{b}}, \qq
\mathcal{L}_2 = - \frac{1}{2} \sqrt{\lambda} h^{ij}
G_{mn} \partial_i X^m \partial_j X^n.
\ee
The generalized light-cone gauge conditions are given by
\bel{GLCC}
X^+ = \kappa \tau, \qq
P_- = - \sqrt{\lambda} h^{0j} G_{-, \mathfrak{a}}
\partial_j X^{\mathfrak{a}} = \sqrt{\lambda} \omega = 
\mathrm{const}.
\ee 
We interpret the second condition as the following relation
for $\dot{X}^-$:
\be
\dot{X}^- = - \left( \frac{h^{01}}{h^{00}} \right) \partial_1 X^-
- \frac{\omega}{G_{--}} - \kappa h^{00}
\left( \frac{G_{+-}}{G_{--}} \right).
\ee
With some work, we have
\be
\mathcal{L}_1 = \mathcal{L}_1' + P_- \dot{X}^-,
\ee
where
\be
\begin{split}
\mathcal{L}_1' &=  \sqrt{\lambda} \kappa \omega 
\left( \frac{G_{+-}}{G_{--}} \right)
+ \frac{1}{2} \sqrt{\lambda} \frac{\omega^2}{h^{00} G_{--}} \cr
&- \frac{1}{2} \sqrt{\lambda} h^{00} \frac{\kappa^2}{G^{++}}
+ \frac{1}{2} \sqrt{\lambda} \frac{G_{--}}{h^{00}} ( \partial_1 X^-)^2
+ \sqrt{\lambda} \omega 
\left( \frac{h^{01}}{h^{00}} \right) \partial_1 X^-.
\end{split}
\ee
Note that $P_- \dot{X}^-$ is a total $\tau$-derivative term.
So, we use $\mathcal{L}_1'$ as the Lagrangian 
in the generalized light-cone gauge.

In $\mathcal{L}_1'$, the field $X^-$ appears 
only through the form of $\partial_1 X^-$.
The field $\partial_1 X^-$ plays the role of an auxiliary field.
The equations of motion for $\partial_1 X^-$ gives
\be
\partial_1 X^- = - \frac{\omega h^{01}}{G_{--}}.
\ee
By substituting this solution into $\mathcal{L}_1'$, we have
\be
\mathcal{L}_1' =  \sqrt{\lambda} \kappa \omega
\left( \frac{G_{++}}{G_{--}} \right)
- \frac{1}{2} \sqrt{\lambda} h^{00} \frac{\kappa^2}{G^{++}}
- \frac{1}{2} \sqrt{\lambda} h^{11} \frac{\omega^2}{G_{--}}.
\ee
Let us introduce a world-sheet symmetric tensor $\mathcal{J}_{ij}$ by
\be
\mathcal{J}_{00}:= \frac{\kappa^2}{G^{++}}, \qq
\mathcal{J}_{11}:= \frac{\omega^2}{G_{--}}, \qq
\mathcal{J}_{01}=\mathcal{J}_{10}:=0.
\ee
The reduced action now has the form 
\be
\begin{split}
\mathcal{L}' &= \mathcal{L}_1' + \mathcal{L}_2 \cr
&=  \sqrt{\lambda} \kappa \omega \left( \frac{G_{+-}}{G_{--}} \right)
- \frac{1}{2} \sqrt{\lambda} h^{ij} ( \mathcal{J}_{ij}
+ \mathcal{G}_{ij} ),
\end{split}
\ee
where
\be
\mathcal{G}_{ij} = G_{mn} \partial_i X^m \partial_j X^n.
\ee

Since the world-sheet diffeomorphism is fixed 
by the light-cone gauge conditions \eqref{GLCC}, $h^{ij}$ are determined by
solving the equations of motion for $h^{ij}$:
\be
h^{ij} = \pm \sqrt{- \det( J_{ij})}\, J^{ij},
\ee
where $J_{ij} = \mathcal{J}_{ij} + \mathcal{G}_{ij}$, 
and $J^{ij}$ is the inverse of $J_{ij}$.

Then, we finally have the reduced Lagrangian
in the generalized light-cone gauge
\be
\mathcal{L}'
= \pm \sqrt{ - \lambda \det( \mathcal{J}_{ij} + \mathcal{G}_{ij} )}
+ \sqrt{\lambda}\kappa \omega 
\left( \frac{G_{+-}}{G_{--}} \right).
\ee
 
\section{The GS action}

Now let us consider the GS action
in the $AdS_5 \times S^5$ background. 
The GS action in the flat target space
\cite{GS,GS2} is generalized in the curved supergravity background in \cite{GHMNT}.
More explicit GS action in the $AdS_5 \times S^5$ background was constructed in 
\cite{MT}. (See also \cite{KRR,MT2}).
Originally, the Wess-Zumino term is written in the
three-dimensional form. 
The manifestly two-dimensional
form of the Wess-Zumino term was presented in \cite{BBHZZ,ber,RS}.

We write the GS action in the $AdS_5 \times S^5$ background
as follows:
\be
S_{\mathrm{GS}} = \frac{1}{2\pi} \int \de^2 \xi \, \mathcal{L}_{\mathrm{GS}},
\ee
\be
\mathcal{L}_{\mathrm{GS}}
= - \frac{1}{2} \sqrt{\lambda} h^{ij} \eta_{\ul{a}\, \ul{b}}
E^{\ul{a}}_i E^{\ul{b}}_j
+ \sqrt{\lambda} \epsilon^{ij} ( E_i^{\ul{\alpha}}
\varrho_{\ul{\alpha} \ul{\beta}} E^{\ul{\beta}}_j
- \overline{E}^{\ul{\bar{\alpha}}}_i \varrho_{\ul{\bar{\alpha}} 
\ul{\bar{\beta}}}
\overline{E}^{\ul{\bar{\beta}}}_j ).
\ee
Here $\ul{a}, \ul{b}=0,1,\dotsc, 9$, $\ul{\alpha}, \ul{\beta},
\ul{\bar{\alpha}}, \ul{\bar{\beta}} = 1,2,\dotsc, 16$,
$h^{ij} = \sqrt{-g} g^{ij}$ ($i,j=0,1$), $\epsilon^{01}=1$.
\be
\eta_{\ul{a}\, \ul{b}} = \mathrm{diag}(-1,1,\dotsc, 1).
\ee
The induced vielbein for the type IIB superspace $E^A{}_i$
is denoted by
\be
E^A_i = E^A{}_M \partial_i Z^M
= E^A{}_{\ul{m}} \partial_i X^{\ul{m}}
+ E^A{}_{\ul{\alpha}}\partial_i \theta^{\ul{\alpha}}
+ \ol{E}^A{}_{\ul{\bar{\alpha}}}
\partial_i \bar{\theta}^{\ul{\bar{\alpha}}}, \qq
A= ( \ul{a}, \ul{\alpha}, \ul{\bar{\alpha}} ).
\ee
We use a Majorana-Weyl representation for the Gamma matrices:
\be
\Gamma^{\ul{a}} = \mat{ 0 & \ & 
( \gamma^{\ul{a}} )^{\ul{\alpha} \ul{\beta}} \cr
( \gamma^{\ul{a}} )_{\ul{\alpha} \ul{\beta}} & & 0 },
\qq
( \Gamma^{\ul{a}} )^* = \Gamma^{\ul{a}},
\qq \{ \Gamma^{\ul{a}}, \Gamma^{\ul{b}} \} = 2 \eta^{\ul{a}\, \ul{b}}
1_{32},
\ee
$\ul{a}=0,1,\dotsc, 9$, 
$\ul{\alpha}, \ul{\beta} = 1,2,\dotsc, 16$.
We denote the $n \times n$ identity matrix by $1_n$.
For our specific choice of the Gamma matrices, see appendix.

A $32$-component Weyl spinor $\Theta$ with positive chirality 
has the following form in the Majorana-Weyl representation:
\be
\Theta = \mat{ \theta^{\ul{\alpha}} \cr 0 }.
\ee
Above, we have used the $16$-component notation for the Weyl spinors.
A spinor with upper index $\ul{\alpha}$ represent a Weyl spinor with positive chirality.

The constant matrix $\varrho$ in the Wess-Zumino term is given by
\be
C \Gamma^{01234} = \mat{ \varrho_{\ul{\alpha}\, \ul{\beta}} & \ & 0 \cr
0 & & \varrho^{\ul{\alpha}\, \ul{\beta}} }.
\ee
Here $C$ is the charge conjugation matrix.

\subsection{$\kappa$-symmetry fixing}

Let us decompose each of the two $16$-component Weyl spinors into two $8$-component
$SO(4) \times SO(4)$ spinors:
\be
\theta^{\ul{\alpha}} = \mat{ \theta^{+\alpha} \cr \theta^{-\dot{\alpha}} }, \qq
\bar{\theta}^{\ul{\bar{\alpha}}}
= \mat{ \bar{\theta}^{+\bar{\alpha}} \cr \bar{\theta}^{-\dot{\bar{\alpha}}} },
\ee
where $\alpha=1,2,\dotsc, 8$, $\dot{\alpha} = \dot{1}, \dot{2}, \dotsc, \dot{8}$,
$\bar{\alpha}=\bar{1}, \bar{2}, \dotsc, \bar{8}$ and
$\dot{\bar{\alpha}} = \dot{\bar{1}}, \dot{\bar{2}}, \dotsc, \dot{\bar{8}}$.

We first fix the $\kappa$-symmetry by setting 
$\theta^{-\dot{\alpha}} = \bar{\theta}^{-\dot{\bar{\alpha}}} = 0$.
In the $32$-component notation, these conditions are equivalent to the condition
$\Gamma^+ \Theta = 0$.

To simplify expressions, we combine the remaining fermionic coordinates 
into $\Psi^{\hat{\alpha}}$:
\be
( \Psi^{\hat{\alpha}} ) = \mat{ \theta^{+\alpha} \cr \bar{\theta}^{+\bar{\alpha}} }, \qq
\hat{\alpha} = \hat{1}, \hat{2}, \dots, \hat{16}.
\ee

Let $\mathcal{M}^2$ be a $16 \times 16$ matrix
\be
\mathcal{M}^2 = \mat{
(\mathcal{M}^2)^{\alpha}{}_{\beta} & \ &
(\mathcal{M}^2)^{\alpha}{}_{\bar{\beta}} \cr
(\mathcal{M}^2)^{\bar{\alpha}}{}_{\beta} & \ &
(\mathcal{M}^2)^{\bar{\alpha}}{}_{\bar{\beta}} },
\ee
with elements constructed purely from the fermionic variables:
\be
\begin{split}
( \mathcal{M}^2)^{\alpha}{}_{\beta}
&= \frac{1}{2} ( \theta^+ \gamma_{ab} )^{\alpha}
( \bar{\theta}^+ \gamma^{ab} \varrho)_{\beta}
- \frac{1}{2} ( \theta^+ \gamma_{a'b'} )^{\alpha}
( \bar{\theta}^+ \gamma^{a'b'} \varrho)_{\beta}, \cr
( \mathcal{M}^2)^{\alpha}{}_{\bar{\beta}}
&= - \frac{1}{2} ( \theta^+ \gamma_{ab} )^{\alpha}
( \theta^+ \gamma^{ab} \varrho)_{\bar{\beta}}
+ \frac{1}{2} ( \theta^+ \gamma_{a'b'} )^{\alpha}
( \theta^+ \gamma^{a'b'} \varrho)_{\bar{\beta}}, \cr
( \mathcal{M}^2)^{\bar{\alpha}}{}_{\beta}
&= \frac{1}{2} ( \bar{\theta}^+ \gamma_{ab} )^{\bar{\alpha}}
( \bar{\theta}^+ \gamma^{ab} \varrho)_{\beta}
- \frac{1}{2} ( \bar{\theta}^+ \gamma_{a'b'} )^{\bar{\alpha}}
( \bar{\theta}^+ \gamma^{a'b'} \varrho)_{\beta}, \cr
( \mathcal{M}^2)^{\bar{\alpha}}{}_{\bar{\beta}}
&= - \frac{1}{2} ( \bar{\theta}^+ \gamma_{ab} )^{\bar{\alpha}}
( \theta^+ \gamma^{ab} \varrho)_{\bar{\beta}}
+ \frac{1}{2} ( \bar{\theta}^+ \gamma_{a'b'} )^{\bar{\alpha}}
( \theta^+ \gamma^{a'b'} \varrho)_{\bar{\beta}}.
\end{split}
\ee
Here $a,b=1,2,3,4$, $a',b'=5,6,7,8$.

For later convenience, let us introduce the following matrices:
\be
\frac{\cosh \mathcal{M}-1_{16}}{\mathcal{M}^2}
= \mat{ (K_{11})^{\alpha}{}_{\beta} & \ & 
(K_{12})^{\alpha}{}_{\bar{\beta}} \cr
(K_{21})^{\bar{\alpha}}{}_{\beta} & &
(K_{22})^{\bar{\alpha}}{}_{\bar{\beta}} },
\qq
\frac{\sinh \mathcal{M}}{\mathcal{M}}
= \mat{ (L_{11})^{\alpha}{}_{\beta} & \ & 
(L_{12})^{\alpha}{}_{\bar{\beta}} \cr
(L_{21})^{\bar{\alpha}}{}_{\beta} & &
(L_{22})^{\bar{\alpha}}{}_{\bar{\beta}} }.
\ee

The $\kappa$-symmetry fixed action in the 
$AdS_5 \times S^5$ can be written as \cite{IO}:
\bel{KLGS}
\mathcal{L}_{\mathrm{GS}}
= - \frac{1}{2} \sqrt{\lambda} h^{ij} G_{\ul{m}\, \ul{n}}
\mathcal{D}_i X^{\ul{m}} \mathcal{D}_j X^{\ul{n}}
+ \frac{1}{2} \sqrt{\lambda} \epsilon^{ij}
B_{\hat{\alpha} \hat{\beta}} \mathcal{D}_i \Psi^{\hat{\alpha}}
\mathcal{D}_j \Psi^{\hat{\beta}}.
\ee
Here $\ul{m}, \ul{n}=0,1,\dotsc, 9$, $\hat{\alpha}, \hat{\beta}
= \hat{1},\hat{2},\dotsc, \hat{16}$. The target space metric $G_{\ul{m}\, \ul{n}}$
is the bosonic $AdS_5 \times S^5$ metric, chosen as follows:
\be
\de s^2 = \de s^2_{AdS_5} + \de s^2_{S^5} = 
G_{\ul{m}\, \ul{n}} \de X^{\ul{m}} \de X^{\ul{n}}
= G_{\mathfrak{a} \mathfrak{b}} \de X^{\mathfrak{a}} \de X^{\mathfrak{b}}
+ G_{mn} \de X^m \de X^n.
\ee
The $AdS_5$ part metric is chosen as
\be
\de s^2_{AdS_5} = - \left( \frac{1 + (z^2/4)}{1-(z^2/4)} \right)^2 \de t^2
+ G_z \sum_{a=1}^4 (\de z^a)^2,
\ee
and the $S^5$ part metric is chosen as
\be
\de s^2_{S^5} = \left( \frac{1-(y^2/4)}{1+(y^2/4)} \right)^2 \de \varphi^2
+ G_y \sum_{s=1}^4 ( \de y^s)^2,
\ee
where
\be
G_z = \frac{1}{(1-(z^2/4))^2}, \qq
G_y = \frac{1}{(1+(y^2/4))^2}.
\ee
Here
\be
z^2 = \sum_{a=1}^4 (z^a)^2, \qq
y^2 = \sum_{s=1}^4 (y^s)^2.
\ee
We choose the coordinates $X^{\ul{m}}$ as follows:
\be
X^{\pm} = \frac{1}{\sqrt{2}} ( t \pm \varphi), \qq
X^a = z^a, \qq X^{4+s} = y^s.
\ee
The metric for the longitudinal directions 
$G_{\mathfrak{a} \mathfrak{b}}$ $(\mathfrak{a}, \mathfrak{b}=\pm$) is given by 
\be
G_{++} = G_{--} = - \frac{1}{2} \left( \frac{1+(z^2/4)}{1-(z^2/4)} \right)^2
+ \frac{1}{2} \left( \frac{1 - (y^2/4)}{1+(y^2/4)} \right)^2, 
\ee
\be
G_{+-} = G_{-+} = - \frac{1}{2} \left( \frac{1+(z^2/4)}{1-(z^2/4)} \right)^2
- \frac{1}{2} \left( \frac{1 - (y^2/4)}{1+(y^2/4)} \right)^2. 
\ee
For $G_{mn}$ $(m,n=1,2,\dotsc, 8)$, we have
\be
G_{ab} = G_z \delta_{ab}, \qq
G_{4+s,4+s'} = G_y \delta_{s,s'}, \qq
G_{a,4+s} = 0.
\ee
Here $a,b=1,2,3,4$, $s,s'=1,2,3,4$. Let us denote the inverse of $G_{\ul{m}\, \ul{n}}$
by $G^{\ul{m}\, \ul{n}}$.
Note that $\partial/\partial X^{\pm}$ are Killing vectors.

The derivatives $\mathcal{D}_j$ in \eqref{KLGS} are given by
\be
\begin{split}
\mathcal{D}_i X^+ &= \partial_i X^+, \cr
\mathcal{D}_i X^- &= \partial_i X^- 
+ \Lambda^-{}_{\hat{\alpha}} \mathcal{D}_i \Psi^{\hat{\alpha}}, \cr
\mathcal{D}_i X^m &= \partial_i X^m + ( \Lambda^m{}_{n \hat{\alpha}}
\mathcal{D}_i \Psi^{\hat{\alpha}}) X^n, \cr
\mathcal{D}_i \Psi^{\hat{\alpha}} &= \partial_i \Psi^{\hat{\alpha}}
+ ( \Lambda^{\hat{\alpha}}{}_{\hat{\beta}} \partial_i X^+) \Psi^{\hat{\beta}}.
\end{split}
\ee
$\hat{\alpha} = ( \alpha, \bar{\alpha})$, $\hat{\alpha}=1,2,\dotsc, 16$,
$\alpha, \bar{\alpha}=1,2, \dotsc, 8$. 
The terms $\Lambda^-{}_{\hat{\alpha}}$ in $\mathcal{D}_i X^-$ are given by
\be
\begin{split}
\Lambda^-{}_{\alpha} &= 2 \sqrt{2} \im 
\Bigl[ ( \bar{\theta}^+ \gamma_+ K_{11} )_{\alpha}
+ ( \theta^+ \gamma_+ K_{21})_{\alpha} \Bigr],  \cr
\Lambda^-{}_{\bar{\alpha}} &= 2 \sqrt{2} \im
\Bigl[ ( \bar{\theta}^+ \gamma_+ K_{12} )_{\bar{\alpha}}
+ ( \theta^+ \gamma_+ K_{22})_{\bar{\alpha}} \Bigr], 
\end{split}
\ee
where
\be
( \gamma_+)_{\ul{\alpha}\, \ul{\beta}} = \frac{1}{2} 
\Bigl( ( \gamma_0)_{\ul{\alpha}\, \ul{\beta}} + ( \gamma_9)_{\ul{\alpha}\, \ul{\beta}}
\Bigr)
= \mat{ 1_8 & \ & 0 \cr 0 & & 0 }.
\ee
The terms $\Lambda^m{}_{n\hat{\alpha}}$ 
in $\mathcal{D}_i X^m = ( \mathcal{D}_i z^a, \mathcal{D}_i y^s)$ are given by
\be
\begin{split}
\Lambda^a{}_{b\alpha}
&= - 2 \left[ ( \bar{\theta}^+ \gamma^{ab} \varrho K_{11})_{\alpha}
- ( \theta^+ \gamma^{ab} \varrho K_{21} )_{\alpha} \right], \cr
\Lambda^a{}_{b\bar{\alpha}}
&= - 2 \left[ ( \bar{\theta}^+ \gamma^{ab} \varrho K_{12})_{\bar{\alpha}}
- ( \theta^+ \gamma^{ab} \varrho K_{22} )_{\bar{\alpha}} \right], \cr
\Lambda^a{}_{(4+s) \hat{\alpha}} &= 0,
\end{split}
\ee
\be
\begin{split}
\Lambda^{4+s}{}_{a \hat{\alpha}} &= 0, \cr
\Lambda^{4+s}{}_{(4+s') \alpha}
&= 2 \left[ ( \bar{\theta}^+ \gamma^{4+s,4+s'} \varrho K_{11})_{\alpha}
- ( \theta^+ \gamma^{4+s,4+s'} \varrho K_{21} )_{\alpha} \right],
\cr
\Lambda^{4+s}{}_{(4+s') \bar{\alpha}}
&= 2 \left[ ( \bar{\theta}^+ \gamma^{4+s,4+s'} \varrho K_{12})_{\bar{\alpha}}
- ( \theta^+ \gamma^{4+s,4+s'} \varrho K_{22} )_{\bar{\alpha}} \right].
\end{split}
\ee
The terms $\Lambda^{\hat{\alpha}}{}_{\hat{\beta}}$ in $\mathcal{D}_i \Psi^{\hat{\alpha}}
= ( \mathcal{D}_i \theta^{+\alpha}, \mathcal{D}_i \bar{\theta}^{+\bar{\alpha}})$
are given by
\be
\Lambda^{\alpha}{}_{\beta} = - \frac{\im}{\sqrt{2}} 
( \gamma_+ \varrho)_{\beta}{}^{\alpha}, \qq
\Lambda^{\bar{\alpha}}{}_{\bar{\beta}}
= \frac{\im}{\sqrt{2}} ( \gamma_+ \varrho)_{\bar{\beta}}{}^{\bar{\alpha}}, \qq
\Lambda^{\alpha}{}_{\bar{\beta}} = \Lambda^{\bar{\alpha}}{}_{\beta} = 0.
\ee

The fields $B_{\hat{\alpha}\hat{\beta}}$ in the
Wess-Zumino term in \eqref{KLGS} are defined by
\bel{Bab}
\begin{split}
B_{\alpha \beta} 
&= 2 W_{\gamma \delta}
\left( ( L_{11})^{\gamma}{}_{\alpha} ( L_{11})^{\delta}{}_{\beta}
- ( L_{21})^{\bar{\gamma}}{}_{\alpha} ( L_{21} )^{\bar{\delta}}{}_{\beta} 
\right), \cr
B_{\alpha \bar{\beta}}
&= 2 W_{\gamma \delta}
\left( ( L_{11})^{\gamma}{}_{\alpha} ( L_{12})^{\delta}{}_{\bar{\beta}}
- ( L_{21})^{\bar{\gamma}}{}_{\alpha} ( L_{22} )^{\bar{\delta}}{}_{\bar{\beta}} 
\right), \cr
B_{\bar{\alpha} \beta}
&= 2 W_{\gamma \delta}
\left( ( L_{12})^{\gamma}{}_{\bar{\alpha}} ( L_{11})^{\delta}{}_{\beta}
- ( L_{22})^{\bar{\gamma}}{}_{\bar{\alpha}} ( L_{21} )^{\bar{\delta}}{}_{\beta} 
\right), \cr
B_{\bar{\alpha} \bar{\beta}}
&= 2 W_{\gamma \delta}
\left( ( L_{12})^{\gamma}{}_{\bar{\alpha}} ( L_{12})^{\delta}{}_{\bar{\beta}}
- ( L_{22})^{\bar{\gamma}}{}_{\bar{\alpha}} 
( L_{22} )^{\bar{\delta}}{}_{\bar{\beta}} 
\right),
\end{split}
\ee
where
\be
W_{\alpha \beta}
= \frac{\bigl[
 (1 + (z^2/4))(1 - (y^2/4))  \varrho_{\alpha \beta}
- z^a y^s ( \gamma_a \gamma_{4+s} \varrho )_{\alpha \beta}
\bigr]}
{(1-(z^2/4))(1 + (y^2/4))}.
\ee

\subsection{Generalized light-cone gauge}

Now let us consider the relevant part of \eqref{KLGS}
\be
\mathcal{L}_1:= - \frac{1}{2} \sqrt{\lambda} h^{ij} 
G_{\mathfrak{a}\mathfrak{b}} \mathcal{D}_i X^{\mathfrak{a}} \mathcal{D}_j
X^{\mathfrak{b}}.
\ee
The generalized light-cone gauge is chosen as
\be
X^+ = \kappa \tau, \qq
P_- = - \sqrt{\lambda} h^{0i}G_{-, \mathfrak{a}} \mathcal{D}_i X^{\mathfrak{a}}
=:  \sqrt{\lambda} \omega = \mathrm{const}.
\ee
From the second equation, we have
\be
\mathcal{D}_0 X^- = - \frac{\omega}{h^{00} G_{--}}
- \kappa \left( \frac{G_{+-}}{G_{--}} \right)
- \left( \frac{h^{01}}{h^{00}} \right) \mathcal{D}_1 X^-,
\ee
or
\be
\dot{X}^- = - \Lambda^-{}_{\hat{\alpha}} \mathcal{D}_0 \Psi^{\hat{\alpha}}
- \frac{\omega}{h^{00} G_{--}}
- \kappa \left( \frac{G_{+-}}{G_{--}} \right)
- \left( \frac{h^{01}}{h^{00}} \right) \mathcal{D}_1 X^-.
\ee
\be
\mathcal{L}_1 = \mathcal{L}_1' + P_- \dot{X}^-,
\ee
\be
\begin{split}
\mathcal{L}_1' &=  \sqrt{\lambda} \kappa \omega \left( \frac{G_{+-}}{G_{--}} \right)
+ \sqrt{\lambda} \omega \Lambda^-{}_{\hat{\alpha}}\mathcal{D}_0 \Psi^{\hat{\alpha}} \cr
& - \frac{1}{2} \sqrt{\lambda} h^{00} \frac{\kappa^2}{G^{++}}
+ \frac{1}{2} \sqrt{\lambda}\frac{\omega^2}{h^{00} G_{--}} \cr
& + \frac{1}{2} \sqrt{\lambda} \frac{G_{--}}{h^{00}} ( \mathcal{D}_1 X^- )^2
+ \sqrt{\lambda} \omega \left( \frac{h^{01}}{h^{00}} \right) \mathcal{D}_1 X^-.
\end{split}
\ee
Now $\mathcal{D}_1 X^-$ is an auxiliary field. By solving the equations of motion,
we have
\be
\mathcal{D}_1 X^-= - \frac{\omega h^{01}}{G_{--}}.
\ee
Substitution of this relation into $\mathcal{L}_1'$ yields
\be
\begin{split}
\mathcal{L}_1' &= 
+ \sqrt{\lambda} \kappa \omega \left( \frac{G_{+-}}{G_{--}} \right)
+ \sqrt{\lambda} \omega \Lambda^-{}_{\hat{\alpha}}\mathcal{D}_0 \Psi^{\hat{\alpha}} \cr
& - \frac{1}{2} \sqrt{\lambda} h^{00}
\left( \frac{\kappa^2}{G^{++}} \right)
- \frac{1}{2} \sqrt{\lambda} h^{11} \left( \frac{\omega^2}{G_{--} }\right).
\end{split}
\ee
Let
\be
\mathcal{J}_{00}:= \frac{\kappa^2}{G^{++}}, \qq 
\mathcal{J}_{11} := \frac{\omega^2}{G_{--}}, \qq
\mathcal{J}_{01} = \mathcal{J}_{10}:=0.
\ee

The GS action in the generalized light-cone gauge is given by
\be
\begin{split}
\mathcal{L}_{\mathrm{GS}}' 
&= - \frac{1}{2} \sqrt{\lambda} h^{ij} ( \mathcal{J}_{ij} + \mathcal{G}_{ij} ) 
+ \frac{1}{2} \sqrt{\lambda}
\epsilon^{ij} B_{\hat{\alpha} \hat{\beta}}
\mathcal{D}_i \Psi^{\hat{\alpha}} \mathcal{D}_j \Psi^{\hat{\beta}} \cr
& + \sqrt{\lambda} \kappa \omega \left( \frac{G_{+-}}{G_{--}} \right)
+ \sqrt{\lambda} \omega \Lambda^-{}_{\hat{\alpha}}
\mathcal{D}_0 \Psi^{\hat{\alpha}},
\end{split}
\ee
where
\be
\mathcal{G}_{ij} = G_{mn} \mathcal{D}_i X^m \mathcal{D}_j X^n.
\ee
By removing $h^{ij}$, we finally have the GS Lagrangian in the $AdS_5 \times S^5$
in the generalized light-cone gauge:
\bel{GSinLC}
\begin{split}
\mathcal{L}_{\mathrm{GS}}' 
&= \sqrt{ -\lambda \det ( \mathcal{J}_{ij} + \mathcal{G}_{ij}) } 
+ \frac{1}{2} \sqrt{\lambda}\epsilon^{ij} B_{\hat{\alpha} \hat{\beta}}
\mathcal{D}_i \Psi^{\hat{\alpha}} \mathcal{D}_j \Psi^{\hat{\beta}} \cr
& + \sqrt{\lambda} \kappa \omega \left( \frac{G_{+-}}{G_{--}} \right)
+ \sqrt{\lambda} \omega \Lambda^-{}_{\hat{\alpha}}
\mathcal{D}_0 \Psi^{\hat{\alpha}}.
\end{split}
\ee
Here
\be
\mathcal{D}_i \theta^{+\alpha} = \partial_i \theta^{+\alpha}
- \frac{\im \kappa}{\sqrt{2}} \delta_{i,0} ( \varrho \gamma_+ \theta^+)^{\alpha},
\qq
\mathcal{D}_i \bar{\theta}^{+\bar{\alpha}}
= \partial_i \theta^{+\bar{\alpha}}
+ \frac{\im \kappa}{\sqrt{2}} \delta_{i,0} ( \varrho \gamma_+ \bar{\theta}^+)^{\bar{\alpha}},
\ee
\be
\mathcal{D}_i X^m = \partial_i X^m 
+ ( \Lambda^m{}_{n\hat{\alpha}} \mathcal{D}_i \Psi^{\hat{\alpha}}) X^n.
\ee
This is the main result of this paper. It can be rewritten as follows:
\bel{FLGS}
\begin{split}
\mathcal{L}_{\mathrm{GS}}'
&= \sqrt{ - \lambda \det ( \mathcal{J}_{ij} + \mathcal{G}_{ij})}
+ \sqrt{\lambda} \kappa \omega \left( \frac{G_{+-}}{G_{--}} \right) \cr
& + \sqrt{\lambda} \epsilon^{ij}
\mathcal{D}_i \Psi^{\hat{\alpha}}
\left[ \frac{\sinh \mathcal{M}^T}{\mathcal{M}^T} \, \mathcal{W}\, 
\frac{\sinh \mathcal{M}}{\mathcal{M}} \right]_{\hat{\alpha} \hat{\beta}}
\mathcal{D}_j \Psi^{\hat{\beta}} \cr
& + \im \, 2 \sqrt{2 \lambda} \, \omega 
\Psi^{\hat{\alpha}} ( \gamma_+)_{\hat{\alpha} \hat{\beta}}
\left( \frac{\cosh \mathcal{M} - 1_{16}}{\mathcal{M}^2} \right)_{\hat{\beta} \hat{\gamma}}
\mathcal{D}_0 \Psi^{\hat{\gamma}}.
\end{split}
\ee
Here
\be
\left[ \frac{\sinh \mathcal{M}^T}{\mathcal{M}^T} \, \mathcal{W}\, 
\frac{\sinh \mathcal{M}}{\mathcal{M}} \right]_{\hat{\alpha} \hat{\beta}}
= 
\left. \left[ \frac{\sinh \mathcal{M}^T}{\mathcal{M}^T} \right]_{\hat{\alpha}}
\right.^{\hat{\gamma}}\, 
\mathcal{W}_{\hat{\gamma} \hat{\delta}} \, 
\left.\left[ \frac{\sinh \mathcal{M}}{\mathcal{M}} \right]^{\hat{\delta}}
\right._{\hat{\beta}},
\ee
\be
\left. \left[ \frac{\sinh \mathcal{M}^T}{\mathcal{M}^T} \right]_{\hat{\alpha}}
\right.^{\hat{\gamma}}
= \left.\left[ \frac{\sinh \mathcal{M}}{\mathcal{M}} \right]^{\hat{\gamma}}
\right._{\hat{\alpha}},
\qq
\mathcal{W}_{\hat{\gamma} \hat{\delta}}
= \mat{ W_{\gamma \delta} & \ & 0 \cr
0 & & W_{\bar{\gamma} \bar{\delta}} },
\ee
with $W_{\bar{\gamma} \bar{\delta}} = W_{\gamma \delta}$, and
\be
( \gamma_+)_{\hat{\alpha} \hat{\beta}}
= \mat{ 0 & \ & ( \gamma_+)_{\alpha \bar{\beta}} \cr
( \gamma_+)_{\bar{\alpha} \beta} & & 0 }
= \mat{ 0 & \ & 1_8 \cr 1_8 & & 0 }. 
\ee

\subsection{Flat space limit}

In the last subsection, we show that the Green-Schwarz action
in the generalized light-cone gauge \eqref{GSinLC} has the correct
flat space limit.

In order to have consistent flat space limit, the following assumptions are necessary:
\be
\kappa < 0, \qq \omega > 0.
\ee
Let us rescale the fields as follows:
\be
X^m = \lambda^{-1/4} \tilde{X}^m, \qq m=1,2,\dotsc, 8,
\ee
\be
\theta^{+\alpha} = \frac{\lambda^{-1/8}}{\sqrt{2}}( S_1^{\alpha} + \im S_2^{\alpha}), \qq
\bar{\theta}^{+\bar{\alpha}}
= \frac{\lambda^{-1/8}}{\sqrt{2}} ( S_1^{\alpha} - \im S_2^{\alpha}), \qq
\alpha=1,2,\dotsc, 8.
\ee
We also need to set the parameters as follows:
\be
\kappa = - \lambda^{-1/4} \tilde{\kappa}, \qq
\omega = \lambda^{-1/4} \tilde{\kappa}, \qq
\tilde{\kappa} > 0.
\ee
The flat space limit corresponds to $\lambda \rightarrow \infty$ with
$\tilde{\kappa}$ fixed. 
In the flat space, one of the light-cone gauge condition corresponds to
$\tilde{X}^+ = - \tilde{\kappa} \tau$.

Taking $\lambda \rightarrow \infty$, and
ignoring (divergent) surface terms, the GS Lagrangian in the 
generalized light-cone gauge arrives at
\be
\mathcal{L}'^{\mathrm{flat}}_{\mathrm{GS}}
= \frac{1}{2} \sum_{m=1}^8 \left[ ( \partial_0 \tilde{X}^m)^2 - ( \partial_1 \tilde{X}^m)^2
\right]
+ \sqrt{2} \im \tilde{\kappa}
\sum_{\alpha=1}^8 \Bigl[ S_1^{\alpha}( \partial_0 + \partial_1 ) S_1^{\alpha}
+ S_2^{\alpha}( \partial_0 - \partial_1) S_2^{\alpha} \Bigr].
\ee
This is the correct form of the Green-Schwarz action in the light-cone gauge.

\vspace{0.3cm}

%%%%%%%%%%%%%%%%%%%%%%%%%%%%%%%%%%%%%%%%

\noindent

{\bf{Acknowledgements}}

\vspace{3mm}

This work is supported by the 21 COE program
``Construction of wide-angle mathematical basis focused on knots".
The work of H.I. is supported by the Grant-in-Aid for Scientific
Research (No. 18540285) from Japan Ministry of Education.
The work of T.O. is supported by the Grant-in-Aid for Scientific
Research (No. 18540285 and No. 19540304)
from Japan Ministry of Education.

%%%%%%%%%%%%%%%%%%%%%%%%%%%%%%%%%%%%%%%%%%%%%%%%%%%%%%%%%%%%%%%%%%%%%%%%%%%%%%%%%%

\appendix

\section{Notation}

\subsection{Gamma matrices}

Our choice of the $32 \times 32$ Gamma matrices is given by
\be
\Gamma^{\ul{a}} = \mat{ 0 & \ & ( \gamma^{\ul{a}})^{\ul{\alpha}\, \ul{\beta}} \cr
( \gamma^{\ul{a}} )_{\ul{\alpha}\, \ul{\beta}} & & 0 }, \qq
\{ \Gamma^{\ul{a}}, \Gamma^{\ul{b}} \} = 2 \eta^{\ul{a}\, \ul{b}} 1_{32},
\ee
with
\be
\bigl(( \gamma^{\ul{a}})^{\ul{\alpha}\, \ul{\beta}} \bigr)= ( 1_{16}, \sigma^i ), \qq
\bigl( ( \gamma^{\ul{a}})_{\ul{\alpha}\, \ul{\beta}} \bigr) 
= ( - 1_{16}, \sigma^i ),
\ee
where $\sigma^i$ are real symmetric $16 \times 16$ matrices which satisfy the $SO(9)$
Clifford algebra:
\be
\{ \sigma^i, \sigma^j \} = 2 \delta^{ij} 1_{16}, \qq i,j=1,2,\dotsc, 9.
\ee
The Gamma matrices in this ``Majorana-Weyl'' representation
are real:
\be
( \Gamma^{\ul{a}} )^* =  \Gamma^{\ul{a}}, \qq
\ul{a}=0,1,\dotsc, 9.
\ee
A choice of $\sigma^i$ is \cite{GPvN02d}:
\be
\begin{split}
\sigma^1 &= \tau_2 \otimes \tau_2 \otimes \tau_2 \otimes \tau_2, \cr
\sigma^2 &= \tau_2 \otimes \tau_2 \otimes \tau_1 \otimes 1_2, \cr
\sigma^3 &= \tau_2 \otimes \tau_2 \otimes \tau_3 \otimes 1_2, \cr
\sigma^4 &= \tau_2 \otimes 1_2 \otimes \tau_2 \otimes \tau_1, \cr
\sigma^5 &= \tau_2 \otimes 1_2 \otimes \tau_2 \otimes \tau_3, \cr
\sigma^6 &= \tau_2 \otimes \tau_1 \otimes 1_2 \otimes \tau_2, \cr
\sigma^7 &= \tau_2 \otimes \tau_3 \otimes 1_2 \otimes \tau_2, \cr
\sigma^8 &= \tau_1 \otimes 1_2 \otimes 1_2 \otimes 1_2, \cr
\sigma^9 &= \tau_3 \otimes 1_2 \otimes 1_2 \otimes 1_2.
\end{split}
\ee
Here $\tau_k$ ($k=1,2,3$) are the Pauli matrices.

The charge conjugation matrix $C$ is chosen as
\be
C = \mat{ 0 & \ & 1_{16} \cr -1_{16} & & 0 }.
\ee

The chirality matrix in $D=9+1$ is defined by
\be
\ol{\Gamma}:= \Gamma^0 \Gamma^1 \dotsm \Gamma^9
= \mat{ 1_{16} & \ & 0 \cr 0 & & -1_{16} }.
\ee
Any $32$-components spinor can be written as
\be
\Theta = \mat{ \theta^{\ul{\alpha}} \cr \chi_{\ul{\alpha}} }.
\ee
A Weyl spinor with positive chirality $\ol{\Gamma} \Theta = + \Theta$ is given by
\be
\Theta = \mat{ \theta^{\ul{\alpha}} \cr 0 },
\ee
and a Weyl spinor with negative chirality $\ol{\Gamma} \Theta = - \Theta$ is given by
\be
\Theta = \mat{ 0 \cr \chi_{\ul{\alpha}} }.
\ee
In the $16$-components notation, an spinor $\theta^{\ul{\alpha}}$ ($\chi_{\ul{\alpha}}$)
with the upper (lower) index 
$\ul{\alpha}$ represents a Weyl spinor with positive (negative) chirality.

In this Majorana-Weyl representation, the matrix $\varrho$ is given by
\be
C \Gamma^{012345} = \mat{ \varrho_{\ul{\alpha} \, \ul{\beta}} & \ & 0 \cr
0 & & \varrho^{\ul{\alpha}\,  \ul{\beta}} },
\ee
\be
(\varrho_{\ul{\alpha} \, \ul{\beta}} ) = 1_2 \otimes \tau_2 \otimes \tau_2 \otimes \tau_3,
\qq
( \varrho^{\ul{\alpha} \, \ul{\beta}} )=1_2 \otimes \tau_2 \otimes \tau_2 \otimes \tau_3.
\ee

An antisymmetric product of $\Gamma$'s is denoted by
\be
\Gamma^{\ul{a}\, \ul{b}} = \Gamma^{[\ul{a}} \Gamma^{\ul{b}]}
= \mat{ ( \gamma^{\ul{a}\, \ul{b}})^{\ul{\alpha}}{}_{\ul{\beta}} & \ & 0 \cr
0 & & ( \gamma^{\ul{a}\, \ul{b}})_{\ul{\alpha}}{}^{\ul{\beta}} }.
\ee

\subsection{Indexes}

We uses the following conventions for the various indexes in $AdS_5 \times S^5$.

\vspace{5mm}

\noindent
(i) The world sheet index: $i,j=0,1.$
\[
\xi^0 = \tau, \qq \xi^1 = \sigma, \qq \epsilon^{01} = 1.
\]

\vspace{5mm}

\noindent
(ii) The curved bosonic index : $\ul{m}, \ul{n}=0,1,\dotsc, 9$.

$\ul{m} = ( \mathfrak{a}, m)=
(\mathfrak{a}, a, s)$: $\mathfrak{a}= \pm$, $m=1,2,\dotsc, 8$,
$a=1,2,3,4$, $s=1,2,3,4$.
\[
X^{\ul{m}}: \ \ \ X^0 = t, \qq X^9 = \varphi, \qq
X^{\pm} = \frac{1}{\sqrt{2}}( t \pm \varphi), \qq
\qq X^a = z^a, \qq
X^{4+s} = y^s.
\]
Also $a',b'=5,6,7,8$: $X^{a'} = y^{a'-4}$.

\vspace{5mm}

\noindent
(iii) The local Lorentz index: $\ul{a}, \ul{b}=0,1,\dotsc, 9$,
\[
\eta_{\ul{a}\, \ul{b}} = \mathrm{diag}(-1,+1,\dotsc, +1).
\]

\vspace{5mm}

\noindent
(iv) The index for Weyl spinors: $\ul{\alpha}, \ul{\beta}, \ul{\bar{\alpha}}, 
\ul{\bar{\beta}}=1,2,\dotsc, 16$. 

The bar over $\ul{\alpha}, \ul{\beta}$ has no physical 
meaning, which are occasionally used to distinguish complex Weyl spinors 
from their conjugate:
\[
\theta^{\ul{\alpha}} = \frac{1}{\sqrt{2}}( \theta^{1\ul{\alpha}} + \im
\theta^{2 \ul{\alpha}}), \qq
\bar{\theta}^{\ul{\bar{\alpha}}} = \frac{1}{\sqrt{2}}
( \theta^{1 \ul{\alpha}} - \im \theta^{2 \ul{\alpha}}).
\]
Here $\theta^{I\ul{\alpha}}$ ($I=1,2$) are Majorana-Weyl spinors: 
$( \theta^{I\ul{\alpha}})^* =  \theta^{I \ul{\alpha}}$.

The decomposition of Weyl spinors:
\[
\theta^{\ul{\alpha}} = \mat{ \theta^{+\alpha} \cr \theta^{-\dot{\alpha}} }, \qq
\bar{\theta}^{\ul{\bar{\alpha}}} = \mat{ \bar{\theta}^{+\bar{\alpha}} \cr
\bar{\theta}^{-\dot{\bar{\alpha}}} },
\]
$\ul{\alpha} = ( \alpha, \dot{\alpha})$, $\alpha = 1,2,\dotsc, 8$,
$\dot{\alpha} = \dot{1}, \dot{2}, \dotsc, \dot{8}$.

\vspace{5mm}

\noindent
(v) The index for combined spinors: $\hat{\alpha}, \hat{\beta} = 1,2,\dotsc, 16$.
\be
(\Psi^{\hat{\alpha}}) = \mat{ \theta^{+\alpha} \cr \bar{\theta}^{+\bar{\alpha}} }.
\ee

\end{document}